# The family resemblance of technologically mediated work practices

Eric Monteiro*, Gasparas Jarulaitis*, and Vidar Hepsø**

*Dept. of Computer and information science, Norwegian Univ. of Science and Technology (NTNU), Norway, {gasparas|ericm}@idi.ntnu.no

**Dept. of Petroleum engineering and applied geophysics, Norwegian Univ. of Science and Technology, Norway, vidar.hepso@ntnu.no

**Abstract**

Practice-based perspectives in information systems have established how, in every instance of use (i.e., work practices), the user exercises considerable discretion in their appropriation of the technology with local workarounds and situated improvisations. We analyse the relationship *between* technologically mediated work practices separated in time and space. Specifically, we analyse *how* similarity in work practices is achieved. Achieving absolutely similar (or 'best') practices is unattainable. Drawing on a longitudinal (2007 – 2011) case of ambulatory maintenance work in the oil and gas sector, we identify and discuss three *constituting strategies* (differentiation, assembling and punctuation) through which a *family resemblance* of – similar but not the same – work practices is crafted. We discuss how, in the absence of an essentialist criterion, similarity is subject to pragmatic but also political negotiations.

## 1. Introduction

Users exercise considerable discretion when appropriating technology. Empirical studies have consistently and convincingly demonstrated how information systems routinely, arguably even necessarily, are subject to workarounds, improvisations and tinkering (Gasser, 1986). That every instance of users' interaction with technology is enacted (Orlikowski, 2000), situated (Suchman, 2007), a bricolage (Orr 1996), improvised (Orlikowski, 1996) and contextual (Robey & Sahay, 1996) is well rehearsed in information systems research.

There is, however, a relative scarcity of accounts of the relationship *between* time-space separated *instances* of user's interactions (i.e. work practices). In particular, we lack a robust understanding of how the 'same' work practices emerge over time and space (Leonardi & Barley 2008). In an increasingly globalised world, establishing uniform work practices is vital for competitiveness of business organisations (Leidner 1993). The group of engineers engaged in the maintenance of oil and gas wells we study is under mounting pressure for reasons of economy of scale as well as health, environment and safety improvements to establish more uniform work processes across the many wells they serve.

Quite a few, in line with Garfinkel's (1967) advice, have addressed the time dimension of this relationship: work practices are analysed through processes of learning (Hyysalo 2009; Chu & Robey, 2008), organisational routines (Feldman and Pentland 2003) or imitation/ isomorphism (Batenburg et al. 2008). Fewer studies, notably of Enterprise Systems, also exist that focus on the space dimension and how technology mediates (or rather not) the 'same' ('best') practices across distributed sites (Wagner and Newell 2004; Volkoff et al. 2005; Wagner, Scott, & Galliers, 2006). The purpose of this paper is to contribute towards a theoretical understanding of *how* 'same' technologically mediated work practices emerge by tracing out the space and time dimensions. We make two contributions.

First, we give a *characterisation* of how the 'same' technologically mediated, space-separated work practices emerge over time. As "[t]he vector of time has long been recognised" while "[t]he vector of space, in contrast, has remained comparatively undertheorized" (Amin and Cohendet 2004, p. 86), our analysis starts from technological mediation across space working as a template or plan that over time gets refined, backed up – or defeated. We identify and discuss *constituting strategies* of (i) differentiation ("what is the unique personality of this oil well?"), (ii) assembling similarities ("how to make simplifications?") and (iii) punctuation ("what happens when simplifications break down?").

Second, we discuss *how* similar technologically mediated work practices need to be to be the 'same'. Time-space separated work practices clearly cannot literally be the same. We understand 'similar' not as 'same' (i.e. identical) but as 'family

resemblance[1]' (i.e. degree of sameness). The question thus becomes for whom, where and when are work practices similar *enough*? Family resemblance among work practices is in our analysis not an essentialist quest for certain attributes but rather a performed achievement. We discuss how similar enough is relative to a given purpose framed within political and institutional discourses.

# 2. Theory: Practice-based perspectives and beyond

## 2.1. The local

Practice-based perspectives[2] in information systems do not represent a well-defined body of literature but comprise a loosely connected set of theoretical and methodological approaches (Schatzki et al. 2001; Nicolini 2011). It has been robustly established that technology but create the conditions for – not govern – human encounters with technology (Avgerou and Ciborra, 2004; Boudreau and Robey, 2005; Newell and Wagner, 2006; Orlikowski, 1996, 2000; Robey and Boudreau, 1999). A distinguishing feature of practice-based perspectives is their emphasis on users' interactions with technology (i.e. work practices) as *local* (alternatively: situated, contextual, improvised or enacted). The exact formulation of the affinity with the local varies with the theoretical underpinning. Suchman (2007, p. 70), leaning on a combination of ethnomethodology and science studies, points out that work practice is not predetermined by formal specifications, but instead is contingent and "depends in essential ways on its material and social circumstances". Similarly, from a structuration theory basis, Orlikowski (2000, p.412) underscores the situated, contextual and local nature of a user's interaction as "every encounter with technology is temporally and contextually provisional, and thus there is, in every use, always the possibility of a different structure being enacted".

---

[1]The notion of family resemblance was made famous by Wittgenstein (1953) in his discussion about how we learn the rules of (language) games. There cannot be exhaustive rules telling you when to employ a rule due to the problem of infinite regress. Hence you learn to recognise when similar (i.e. family resemblance) conditions for rules apply.

[2] We use the term broadly to cover information systems research where the contextual conditions for work practices are highlighted, e.g. with notions such as appropriation, practice lens, improvisation, technology-in-practice and enactment (see e.g. Leonardi and Barley (2010) for an extensive review).

Underscoring of the local in practice-based research has entailed an emphasis on tracing out local contingencies at play in information systems implementation processes (Leonardi & Barley, 2010). For example, Robey and Sahay (1996) conducted a comparative case study of a geographical information system (GIS) implementation within two government organisations and identified "radically different experiences with, and consequences of, the GIS technology" (ibid., 93). The difference in outcomes of implementation of the same technology was attributed to differences in the local contexts of the two organisations, e.g. the differences in user involvement, management commitment and technological experience.

Another connotation of the local in practice-based perspectives is the malleable character of the technology. From the field of science studies, technology allows 'interpretive flexibility', implying that "for different social groups, the artefact presents itself as essentially different artefacts" (Bijker, 1992, p.76). Exploring the malleability of technology has been important also in practice-based perspectives in information systems. What practice-based perspectives accomplish well, then, is documenting the inherent space for users' workarounds; there is always leeway for human discretion in encounters with technology (Gasser 1986). The 'agentic turn' (Emirbayer and Mische 1998), which is also influential in practice-based research (Boudreau and Robey 2005), risks privileging individual over collective action, thereby undermining how practice theory originally was geared towards collective practices (Schatzki et al,. 2001; Bourdieu 1977). It is not that practice-based perspectives rule out limitations to local discretion as e.g. Orlikowski (2000, p. 409) writes, "[s]aying that use is situated and not confined to predefined options does not mean that is totally open to any and all possibilities". For instance, Boudreau and Robey (2005) explicitly set out to test whether there is room for human discretion in the case of integrated systems and conclude that their results strengthen the practice-based position by "showing that such enactment apply [also] to an ostensibly less flexible technology, an ERP system" (ibid., p. 14).

The meaning of the 'local' in practice-based research has always been a source of misunderstanding. As Nicolini (2011, p. 603) points out, "the practice-based approach is always exposed to the risk of being understood as a call for more close-up micro-studies". The risk, in other words, is that the local is understood too literally as a space-time confined physical location. An overly literal meaning of 'local' would run counter to a fuller understanding of practice-based perspectives (see e.g. Schatzki et al. 2001; Gherardi 2000). In sum, *that* users' work practices are local/ situated is thus broadly accepted, but exactly *what* a local situation is – how it extends in space and time, how it is mediated by technology – remains contested.

## 2.2. Non-local: enter space and time

Work practices are local in the sense of being shaped by local social, historical and material circumstances but *not* local in the sense of being confined in time-space to a particular locale. Conflating the latter with the former creates seemingly 'contradictions' (Pollock et al. 2009, p. 255), 'puzzles' (Yamauchi and Swanson 2010, p. 188) or as Shapin (1995, p. 307) early put it in the context of the 'artefact' of a scientific fact: "If, as empirical research securely establishes, science is a local product, how does it travel with what seems to be unique efficiency?". How to resolve this contradiction is the subject of ongoing debates in and around information systems research as we outline in discussing three particularly relevant approaches.

One approach is based on insights about standardised (i.e. similar) work practices within science studies. The key is to *embed* the local into the very notion of standardised practice. For instance, Timmermans and Berg (1997) studied the use of clinical protocols (practices). While the primary aim of the protocols is to transform and standardise work practices, the authors vividly illustrate that the protocols are not universal. Actors perform minor and not-so-minor deviations in order to adjust a given protocol to unforeseen situations. Local tinkering with the protocol, the authors argue, is not a failing of but a prerequisite for the working of the protocol. Local tinkering is inherent as captured by their notion of 'local universalism'. As Berg and Timmermans (2000, p. 45) argue, the local tinkering is inherently implied or embedded as the effort itself of standardisation "*produce* the very disorder [local tinkering] they attempt to eradicate".

Within information systems research, Vaast and Walsham (2009) are among the few scholars who have tried to draw on the notion of local universalism (see also Monteiro and Rolland 2012). Vaast and Walsham (2009) studied distributed communities of practice in the field of Environmental Health. The authors emphasised the role of technology, and coined the term 'trans-situated learning' to explain how people can communicate and exchange experience with the help of technology, yet do not share an actual context of work (i.e. separated by a geographical boundary). Pollock et al.'s (2007) notion of 'generification' may like local universalism be understood as an attempt to embed the local within the definition of the universal. They are helpful in dismantling local/global dichotomies, but offer less in terms of detailing the process of crafting local universalism, which brings us to the next approach.

A second approach to resolving the above contradiction is that of re-working the definition/ concept of a 'situation'. Explicitly drawing on, but going beyond Orr's (1996) classic study of repair work of photocopiers, scholars have focused on work practices less reliant on "[t]erritorial boundaries provid[ing] a reference for all action" (Orr 1996, p. 64; also quoted in Pollock et al. 2009, p. 256). From an Activity theory basis, Nicolini (2007) proposes the notion of 'stretching out' the situation in time and space. Central to his analysis is the material mediation of work practices. He concludes that this mediation "implies much more than simple redistribution of existing work" (ibid., p. 914). Driven by related instincts, Pollock et al. (2009) suggest the notion of an 'extended situation'. Focusing on remote problem solving at a help desk of an international software vendor, they analyse how, where and when selected problems were mediated or transformed. The focus in their analysis is to demonstrate the many links to resources, people and routines that make problem solving non-local.

Nicolini (2011) recently attempted to re-frame (the situation, now re-named into) the site. Dismissing the literal (i.e. physical, spatio-temporal) meaning of a site up front, he underscores the *interconnected* nature of practices spread in space and time. His Actor-network theory base leads him to focus on the performances that *make* practices similar, in much the same way as Turnbull (2000, p. 41) who points out that "[t]he answers [to the contradictions about 'local'] lie in a variety of social strategies and technical devices that provide for treating instances of knowledge/ practice as similar or equivalent and for making connections, that is in enabling local knowledge/practices to move and to be assembled".

If similarities of practices are indeed crafted (performed) through sociomaterial *strategies*, this pushes us third and lastly to consider proposals for such strategies applicable to our empirical domain of oil and gas. Frodeman (1986) is an interesting case in point as he outlines strategies related to geology (see also Almklov Hepsø 2011). The geological understanding of one site is established by identifying, through a set of strategies, similarities with other sites.

Frodeman (1996: 418-419) describes a heuristic he calls 'visual intelligence' where a set of templates organise sets of marks into a body of significant signs. The tools used both aid geologists' practice and are instrumental in codifying their observations. He indicates three strategies constituting visual intelligence: *contrasts, patterns* and *aberrancies*. First, contrasts signify boundaries, i.e. what is inside or outside? Contrasts are vital to classification but also to grasp the whole geological understanding and elements of the context. Second, in a search for patterns, geologists bring together a set of similarities and differences that imply order. Crafting order through contrast and finding patterns implies repeatedly moving from the details to the whole (cf. hermeneutics). Third, during these movements in perspective and scale, the geologists keep a keen eye out for anomalies or aberrancies. Anomalies are significant for family resemblance because they deviate from the normal patterns and contrasts with an imposed order. They are clues that may challenge or support interpretation. However, anomalies only make sense as a part of a larger order of contrasts and patterns. They provide pragmatic testing of the robustness of an imposed order envisioned through contrasts and patterns.

In what follows, we operationalise the abstractly formulated strategies of Frodeman (1986) to the work practices of oil and gas well maintenance engineers. We analyse the crafting – temporal and fragile – of similar enough work practices by tracing out the interconnections in space-time of extended situations inspired by Actor-network theory.

# 3. Research methods

## 3.1. Case setting, access and selection

Established in the 1970s, the global oil and gas company (OGC, a pseudonym) has grown from a small, regional operator in Northern Europe to a significant energy company, currently employing some 30,000 people with activities in 40 countries across four continents and listed on the New York Stock Exchange. Historically, the units of OGC have been semi-autonomously organised by geography i.e. around the site of the oil field. The OGC has a long history of organising work according to hierarchical models and a strict division of labour. Currently, OGC may be classified as a matrix organisation with business units serving multiple functions. As result, oil and gas production from a given oil and gas field is dependent on a number of different disciplines belonging to different functional units. In addition to internal matrix organising, the OGC is heavily dependent on multiple external vendors and service companies.

Our study focuses on a designated group of engineers doing oil and gas well maintenance (or 'intervention'). The well is drilled deep into the seabed and is largely inaccessible. Well intervention is critically important to mitigate risks to health, environment and commercial interests. Well interventions have traditionally differed significantly from oil field to oil field due to site-specific differences in the geological formations, layout of the production system and level and profile of competence of the offshore workers. As one engineer explained, "every well is unique".

The three authors of this paper have intimate but different knowledge of OGC (Hepsø et al. 2009; Østerlie et al. 2012). This paper draws predominantly on the empirical data collected by the second author who has been studied collaborative work practices in different organisational contexts in OGC since 2007-2011. The first author has conducted a series of interpretative studies of collaboration and infrastructure in OGC over a period of 15 years. The third author has worked almost 20 years as a senior researcher at OGC. He has been involved in research projects about the operation and maintenance of subsea wells. This is relevant to understand the setting, including the historical context of the oil and gas industry in Northern Europe. The authors, especially the first and third, have an extended history of research collaboration.

Motivated by cost-cutting through economy of scale, OGC has for some time attempted to streamline its core business processes, including well intervention, by working out extensive documentation on the sequence, actors involved and required deliverables during well interventions. These attempts of establishing uniform (i.e. similar) work practices have met with but modest success. In a renewed effort, OGC recently (in 2006) established a so-called lightweight well intervention (LWI) group of engineers that plan and supervise well interventions across large numbers of fields and wells. Establishing (more) similar work practices as LWI illustrates, is a notorious source of controversies and conflict within OGC. The motivation for more similar work practices is clearly cost-efficiency but also safety. Recently (January 2012[3]), several OGC unions have publicly argued that 'excessive' levels of uniformity in key work practices would undermine safe operations. Our selection of LWI was theoretically sampled as an illustration of how tensions over uniform vs. localised work practices, beyond economic performance, are tied to risks to human life and the environment (Eisenhardt 1989).

[3] See industry magazine Teknisk Ukeblad, www.tu.no

### 3.2. Data collection

We draw on three types of data collection mainly conducted by the second author: semi-structured interviews, participatory observations and document studies. We have conducted in total 68 interviews (see Table 1 for an overview), each lasting 1-3 hours and predominately transcribed. The first round of interviews were open-ended and aimed at broadly identifying strategic IT visions, implementation activities related to collaborative systems and users' perceptions of technology. Subsequent interviewing targeted specific infrastructural components, work practices and individual's interactions with technology. The technological complexity and purpose of a collaborative infrastructure were discussed with developers, administrators and managers of the infrastructure. We conducted 14 interviews with members of this group. The 23 interviews with engineers and researchers in the R&D department explored technology-mediated, collaborative work in more detail. The 22 interviews in the Oil and gas prodution (OGP) department focused more closely on key subsurface work practices including drilling, well maintenance, production optimisation and process performance. The 9 interviews at LWI highlighted the specificity of the 'light' well interactions.

--- TABLE 1 about here - ---

We conducted participant observations at several sites within OGC. The second author spent 2-3 days a week during 2008 at the R&D department. He was granted access to OGC's internal computer network and intranet. In 2009, he spent 20 full days of observation over 6 months at an OGC department at another site. This was particularly important to understand the highly technical professional language of the reservoir, production and well engineers comprising the core of the subsurface community. Participant observations were interleaved, as is common with our type of interpretative case studies, with informal interviews/ conversations over coffee or in the corridor to clarify issues, pose a question that could not be put during a meeting or get brief feedback on observations. The LWI group is located in the same building as the OGP department making access smooth. 7 of the 20 days of observations at the OGP site were devoted to LWI.

25 pages of field notes were taken during the observations. The field notes recorded, as accurately as practically feasible, selected episodes, exchanges and outbursts during everyday work. Alongside these ‘raw’ empirical data, we maintained notes about question-begging observations, early interpretations or issues we could not understand (which we needed to clarify with the informants later).

We had access to an extensive collection of predominantly electronic but also paper-based documents. These were mainly internal OGC documents on strategies, plans, memos and experience reports related to the collaborative corporate infrastructure. In addition, we analysed the technical descriptions, formal presentations and training materials of various infrastructural components. A number of presentations, governing documents and formal process descriptions related to LWI activities were studied in detail. Finally, we have collected external reports from industry magazines and the media.

## 3.3. Data analysis

In our longitudinal study, data analysis was iterative and ongoing and overlapping with data collection (Boland 2005). The stages of data analysis are blurred but may, to keep the “inherent creative leap” involved in interpretative research of our kind as transparent as possible (Langley 1999, p. 691), structured into three stages.

First, we produced a flow- or process-oriented understanding of the sequence, content and resources involved in the LWI: how it was requested, planned, conducted and evaluated. Documentary data was particularly useful here. OGC has an extensive, critics hold almost bureaucratic, set of ‘Governing documents’ that describe and regulate business processes including well maintenance.

Second, we manually (using colours, annotations, post-it notes) coded transcripts, field notes and documentary data. Our unit of analysis was LWI practices and their relationship rather than actors (Nicolini 2011). Anything but clean slates (Suddaby 2006), coding was influenced by our deep knowledge of OGC as well as our theoretical affinity with practice- and process-oriented perspectives including Actor-network theory (References suppressed for anonymity). Coding was inductive but with important deductive impulses. Coding was in rounds and interleaved with clustering into conceptual categories (see Table 2). Early coding identified how and where LWI worked around shortcomings of the procedures of the Governing documentation. For instance, complaining that the formal procedures were too crude, LWI engineers stressed the importance of the specific "history of the well" which subsequently aggregated into our conceptual category of 'biography' of the well. Partly driven by deductive influences, early coding did not identify when LWI engineers had enough details about the well to proceed. Rather than precise cut-off criteria, our later coding identified strategies employed that worked as heuristic approximations. For instance, we coded instances where LWI engineers relied on the advice of "more experienced" peers. This coding underscored the important of trust in professional networks and aggregated into one of our conceptual categories (cf. Trust in Table 2). Driven by a healthy scepticism to exhaustive planning and what Orr (1996, p. 110) refers to as the limited 'prescriptive ability' of formal documents, our last round of coding focused on breakdown situations i.e., when plans demonstrably are failing. For instance, instances when "the pressure caus[ed] the tubing to collapse" during LWI, despite plans assessing pressure limits, were coded as the conceptual category of unplanned events.

Third, from the derived conceptual categories we developed our interpretative template in the form of three constructs (see first column of Table 2). Given our ambitions of characterising performative strategies for *how* LWI was accomplished, our constructs mirror our effort to condense the richer insights of the working vehicles, the conceptual categories.

--- TABLE 2 about here ----

# 4. Case

## 4.1. The well

The well is the central object around which exploration, drilling, production, maintenance and process performance evolve. It is a deep, largely inaccessible and highly complex object. Wells vary in length (1000 – 5000 meters), direction (previously only vertical, now placed also horizontally), number of well connected to a facility (smaller ones 10-15 wells while larger fields have 70-100 production wells connected) and purpose (injection of gas and/or water to increase reservoir pressure).
Wells are drilled to extract hydrocarbons from a reservoir to the platform where oil and gas are separated and later transported to onshore facilities by tankers or pipelines. A reservoir contains water and sand in addition to hydrocarbons. Reservoirs differ in their location (depth, geological formation), consistency (porosity, permeability), characteristics (temperature and pressure) and size. Deep-sea and high pressure/temperature wells, of which OGC operate several, require special equipment to be installed to ensure flow assurance and prevent leakage as noted by Obama's commission after the Deepwater Horizon blowout in the Gulf of Mexico in 2010[4].

The majority of OGC's, wells are several decades old. Throughout their life-cycle or *biography*[5], data about the well is collected in a variety of *formats* (text, picture, histograms), degree of *formalisation* (from free-text to time-stamped instrument readings), technological *platforms* (the subsurface community has about numerous specialised information systems for storing, manipulating, analysing and visualising the data) and *purpose* (data collected during exploration focuses on minimizing drilling time whereas production is concerned with the location of the well within the reservoir).

The traditional way since the 1970 to produce offshore wells has been from platforms floating or resting on the seabed. Such wells are called 'topside', as wellheads[6] are installed on the platform. Motivated by a combination of lowered operational costs and strengthened abilities to operate in hostile[7] (large depths, high pressure, cold climate) environments, global energy companies have been fiercely engaged in innovations allowing the production/ extraction of hydrocarbons from 'subsea' wells[8]. In contrast to topside wells, subsea wells are completed on the seabed. The wellhead of a subsea well is installed on the sea floor and not on the platform (see Figure 1 for an illustration). The daily production from subsea wells is from remote (onshore or on neighbouring platforms) control centres based on the instrument readings from the subsea installation.

---

4 See www.oilspillcommission.gov/final-report

5 Recognising the long time-spans of technologies by using the analogy of a 'biography' is borrowed from Pollock and Williams (2009).

6 A wellhead is a part of a well, which terminates at the surface where hydrocarbons can be withdrawn. The wellhead consists of multiple devices that operate the well and ensure production control.

7 'Easily' accessible oil has already been located. The remainder is increasingly inaccessible, involving more elaborate and complex subsea technologies. An estimated 25% of the world's remaining oil reserves are in the Arctic with severe environmental and weather challenges.

8 The vision of both oil companies and vendors is to have complete production plants subsea.

---- INSERT FIGURE 1 ABOUT HERE ----

OGC has strategically targeted subsea wells as the present and planned fields are difficult/ costly to run from top-side wells. Investing heavily since the 1980s, OCG is currently among the global pioneers of subsea wells. OGC operates around 500 subsea wells. On the Norwegian Continental shelf, over 60% of OGC's oil and gas production is currently from subsea wells. This percentage is expected to increase in as "there is a gradual transfer from installations projecting above the sea surface to subsea installations" (OGC intranet news, October 2009). The focus of this paper is on subsea wells and their maintenance.

## 4.2. Well maintenance (or 'intervention')

Subsea wells consist of multiple technological components such as sensors, valves, casing, tubing equipment and electronic control units (see Figure 2). These components are in themselves complex, technological devices obviously not infallible. The components of a subsea well are exposed to severe environmental stress. At different rates, they all decay e.g. from corrosion or sand production (see Figure 3 for illustration). An instant increase in sand production may damage sensor within hours or days. The purpose of well interventions is to maintain the technical integrity of the well to mitigate against health, environmental and/or commercial risks and increase production performance. Unplanned shutting down production due to lacking maintenance represent significant economic damage to OGC from lost revenues and large fixed costs (equipment, personnel).

---- INSERT FIGURES 2 and 3 ABOUT HERE ----

Well interventions mainly repair or replace selected components of the subsea well. Wells can fill with sand and needs 'washing' by injecting chemicals into the well at designated pressure. Many of subsea well interventions are due to 'scale'. Scale consists mainly of inorganic salts that have elements of calcium carbonates, barium and strontium sulphates. The production tubing gets clogged from scale that severely hampers the flow of hydrocarbons in the well. Scale typically develops when reservoir formation water (i.e. water contained inside the reservoir) enters the well. When the formation water undergoes changes in pressure and temperature, or where two incompatible fluids are intermingled, either sulphate or carbonate scales may develop. Even relatively new subsea wells may suffer from scale if drilling or completion fluid is incompatible with the formation water. During production, as oil and gas are gradually drained, increased amounts of formation water are produced together with the hydrocarbons and is likely to give scale challenges.

Well intervention also involves replacing or upgrading faulty or outdated components. Well temperature and pressure transmitters tend to have short life cycles. They are, however, expensive to replace when this entails shutting down the well. Instead, well intervention activities rely on specialised logging tools that do not assume a shut well. Logging tools may gather data to compensate for the loss of instrument measurements. Well logs are collections of different data types based on physical measurements performed by tools lowered into the subsea wells, normally in connection with drilling or well operations.

Well interventions for topside and subsea wells require distinct intervention technologies. While topside wells are accessed from platforms, subsea well interventions are remotely conducted from mobile rigs or vessels. The first subsea well interventions were conducted from mobile rigs, but due to costs, OGC has increasingly used vessels. OGC performed its first light well intervention (LWI) in 2000 and has been committed ever since to this technology due to its high cost saving potential: "deploying a special purpose intervention vessel rather than a rig for downhole operations in subsea wells cuts the costs of these jobs by roughly 50 percent" (OGC intranet news, December 2004). While vessels offer significant cost reduction, they do not eliminate the need for mobile rigs as vessels only perform smaller interventions. If a well is damaged during an intervention, a mobile rig probably has to be employed. It is accordingly important to assess the scope (e.g. rig vs. vessel) of an intervention accurately to leverage the potential efficiency gains of LWI.

## 4.3. Light well intervention (LWI)

Light well intervention (LWI) originated from a UK based energy company. OGC operates a significant number of subsea wells and established a dedicated group to plan and supervise LWI. The LWI department of OGC was established in 2006. It is relatively small (about 30 people) and consists of well and subsea engineers, well planning managers, materials coordinator, health and safety engineers, an economist and a technical assistant who all work onshore and are co-located. In addition, 12 well managers participate in onshore meetings, but primary work onboard offshore vessels. The core activities of LWIs are the *planning* and *supervision* of well interventions. Communication with the central control room at the neighbouring offshore platform that operate the subsea wells during production is vital. The processes of shutting down the well before and re-opening after LWI are safety-critical. While LWI is conducted in one well, normal operations may continue for the other wells belonging to the same field.

LWIs are organised in campaigns. OGC is currently operating two vessels covering approximately 500 subsea wells. Interventions thus have to be planned well in advance. Any field within OGC may request the services of LWI. Typically, a production engineers from a given field submits a request based on his/ her assessment of the local situation. LWI operates a planning matrix outlining the number and location of interventions of the present year. Moreover, well interventions are planned in parallel. The local policy of an LWI is that when a vessel leaves the dock to perform an intervention, two additional interventions have to be planned. In case of unexpected situations, a vessel can perform another intervention:

> We [LWI] do not want vessels to be parked in the dock. The vessel that completes an intervention comes to dock, unloads the equipment, and new equipment is loaded and the vessel leaves to perform another intervention. This happens continually the whole year round… if the vessel is parked in the docks, we loose money… (LWI engineer)

The key challenge for LWI is – drawing on relevant knowledge of involved actors, digging into available documentation about the well and its equipment, an understanding of the reservoir, consulting the production engineer triggering the LWI and the control room of the platform – to grasp the particular issues of the well in question and conduct a sufficiently safe planning and execution of the intervention. LWI is premised on the assumption it can be conducted without insight into the full-fledged biography of the well. For LWI engineers the question is how, where and when simplifications in their work practices are made.

# 5. Analysis

## 5.1. Differentiation: identifying the profile of the well

As most safety-critical organisations, OGC has an extensive set of 'Governing documents' regulating its principal business processes such as exploration, drilling, project development, production, – and well interventions. Governing documents lay out the structure of the process as well as required input to and format of output from these processes. One document template is the so-called 'Well Intervention Assignment' outlining the type of intervention planned together with production and reservoir information. During a 'start-up' meeting, the production engineers present the Well Intervention Assignment to the well engineers, thus initiating LWI.

As is by now well-known, formal work descriptions and templates for LWI, to be applied for all wells uniformly, do not of course govern the work practices but rather act as resources, checklists or a point of departure. In the everyday practices of LWI, a primary strategy is to conjure from resources, including but not restricted to those in the templates, a 'profile' or differentiation of the special characteristics of the well in question. We illustrate.

A central part of LWI's effort to grasp the uniqueness of the well – its *personality* – is to re-construct historically the biography of the well. In stark contrast to topside well interventions, LWI have little prior, local knowledge of the well. On the rare occasions they do have prior experience with the well, "we believe the quality improves" as one LWI engineer stated. Engineers with in-depth, historical understanding of the well conduct topside well interventions. This is exactly what LWI needs to re-create, if not in full, at least sufficiently for the intervention. This re-construction takes time: the planning of topside well interventions typically takes less than a week whereas LWI planning takes more than a month.

There is truly an abundance of data constituting the full biography of the well. To illustrate, during the drilling phase alone, more than a thousand documents may be produced for a single well. For subsea wells, the target of LWI, there is in addition real-time data from instrument readings of pressure (see Figure 4), fluid flow rate, temperature, vibration, composition, fluid hold-up and electromagnetic resistance. One subsea well will typically have about 5-10 sensors.

-- FIGURE 4 about here ---

So how and where to start? The LWI templates provide useful cues by requiring any intervention to include drawings (with coordinates) about the drilling of the well, details about subsea equipment installed, diameter of a well in different zones, description of well completion and experience reports. The so-called Final Well Report is a point of departure. Drilling engineers produce it during completion of the drilling project. It describes equipment used during drilling, experiences and challenges encountered. The historical reconstruction of the well, however, gets entangled with the fact that relevant information is distributed across three generations of IT[9] platforms for electronic archiving:

> You have to dig into several archives [electronic and possibly paper-based], which are usually not accessible by everyone. If it is an electronic archive, you need to get access to it, which can take a lot of time… So I have to find a person who has the authority to give access. Nobody has access to everything. (LWI engineer).

Production engineers, regularly the ones requesting LWI, are less than helpful in the historical reconstruction. As one production engineer explains, "If you didn't follow the well from its inception, there is no way you can know where to find the information or what kind of information that is available" and "there are no defaults…you have to ask people". Compounding the challenge of locating relevant information about the well, naming conventions (for oil and gas fields, wells, documents and archiving structure) are historicised and site-specific: "The problem is that we have a complex tree-structure [of folders] and you have to have been working here for years in order to find something". This literally situated (in history, site-specific) quality of information organisation about the well and the lack of understanding that the well has a biography is the reason why search engines are of limited value. OGC has several times tried to use search engines, but never solved the problem, largely non-existent when we Google the net, of deeply historicized data.

Another crucial aspect of differentiation or working out the personality of a well is the details of the complexity of the *configuration* of the technical components of the subsea well. As one LWI engineer explains, "There is no plug & play possibilities [across vendors of subsea components]", implying that also well work-over equipment used during LWI is proprietary for every vendor. Even compatibility across versions of components from the same vendor may be difficult. Knowing the exact configuration of components of the equipment, crucial for LWI, thus involves consulting closely with the network of external vendors and service providers involved in the well. Even though LWI supervises the interventions, domain experts from several companies involved: the vendor of subsea equipment, the pilots operating the remotely operated vehicles, marine vessel crew and representatives from vendors that did the original completion of the well. Frustrated by not being able to find accurate documentation about the subsea equipment, "I call [the external vendors] and inquire whether they have it. I've done this a couple of times and actually obtained the information" on LWI engineer explains. OGC has recently (2012) started upgrading the control module of subsea wells for one oil field "quite uniquely…as they are mounted also to equipment from a competitor"[10].

---

9 These platforms are: shared disk drives in a Microsoft based network, Lotus Notes databases and Sharepoint.

Lacking (even important) information, however, does not necessarily prohibit LWI engineers from performing an intervention. As one LWI engineer explained:

> If I cannot find specific information, then I use what is available and can conclude that it [the gathered information] is good enough [to perform an intervention] (LWI engineer)

LWI engineers plan an intervention not only to identify as many differences as possible, but also to indentify conditions that mean postponing or cancelling the planned LWI:

> The reason why we spend so much time searching for previous experience is to assess whether it is at all possible to conduct a specific intervention. Earlier experience could indicate specific failures that would prohibit us from doing an intervention. Rather than discovering this when we are onboard the vessel, we find this 1 or 2 months before the operation. For instance, recently we discovered that the control system on the x-mas tree [equipment installed on the sea floor, cf. Figure 3] was not compatible with our equipment. (LWI engineer)

Working out a profile for the well, then, always involves sifting, sampling and simplifying from the (overly) rich, full biography of the well.

---

[10] Quoted by the industry magazine, Teknisk Ukeblad, January 2012.

## 5.2. Assembling similarities: patching together a working understanding

Well interventions are usually performed several years, even decades, after the well was drilled and completed. As pointed out, data related to the well accumulate into vast data sets. Multiple subsurface disciplines work with specific aspects of a well and during distinct phases of well's lifecycle. Every discipline in the subsea community[11] have specialised information systems and produce data with *specific purposes* in mind but may later be re-used by other disciplines. The reuse of data outside its initial and intended context of use is often problematic as there are tacit assumptions about how to make sense of the data ("Is a blank a zero or missing data?", "What kind of equipment was used for this measure?"). This collection of specialised information systems operated by the subsea community is constantly evolving, mutating, integrating, with episodical disruptions far beyond the image of systematically organised portfolios of information systems. They consist of numerous, historically layered information systems where new components partly extend, partly substitute and partly superimpose existing ones. LWI relies heavily on data captured, structured, stored and analysed for purposes other than well interventions. A central concern for LWI engineers is to assess the reliability of the information they use, including but not restricted to the sensor-base data, by comparing and contrasting data: "Forget those choke readings, the sensor stopped working months ago" as one production engineer pointed out. Well engineers thus have to triangulate information i.e. compare information from different sources to assess its. Triangulation is a central activity when planning an intervention. It draws on experience:

> I do not have a lot of experience [a person who has three years experience as a well engineer] and the scary thing with [name of the system] is that I do not necessarily identify mistakes. He [referring to a colleague] can identify mistakes because he has worked with wells for 15 years… certain mistakes you can identify… you can identify that some things are not physically possible... for instance, the diameter of two connected pipes cannot be very different… so some mistakes one can identify, but not all. (LWI engineer)

[11] Subsurface community consists of professionals from a variety of disciplines, e.g. geophysicists, geologists, reservoir engineers, well engineers, production engineers and process engineers.

Triangulation is an informal activity and its extent varies. When performing triangulation, LWI engineers learn about trustworthiness of information sources:

> All of us are aware that information in [name of the system] is not always correct. Preferably, it should be double-checked and compared with other sources, for instance, [name of the system]. For example, information about equipment can be slightly wrong… for instance, the wrong diameter…

While certain systems are deemed 'unreliable' and amendable to double-checking, other systems are trusted more. As one LWI engineer explains, "even if we are not 100% sure, we have to trust [name of the system]".

As for information sources, certain individuals or roles are trusted more than others. For instance, in planning a LWI, one engineer needed the completion reports but realised "you have to know the rig [that did the completion], then you can trace who was responsible for completion". Actors engaged daily with the well are considered trustworthy:

> If I lack specific information or I feel uncertain about something I call an operator [in the platform's control room]. If there are certain limitations in the well, the platform knows about them. So I can talk with an operator and ask. They could say for instance that the annulus pressure should not be higher than 50bar… and then I know this [i.e. that the information is correct] because I have talked to a person who works with that well in the platform every day (LWI engineer)

LWI engineers plan interventions for multiple wells. Over time they learn which information source to trust for specific information. In other words, while information needs for well interventions vary, key (or trustworthy) information sources remain the same and the ways in which information is gathered are similar. Finding the same information across several sources is not a problem but an asset as redundancy increases reliability of information.

In short, LWI engineers collaborate intensively with production engineers and members of the subsurface community in OGC ("I always call the control room [of the platform]"). Collaboration with external vendors and service providers is equally important. Onshore LWI engineers have regular meetings with LWI engineers onboard the vessels. Equipment vendors are involved in the planning process in order to deliver or manufacture certain equipment, or provide reliable information about the exact configuration of the equipment. Close collaboration with vendors is required in order to make sure that a vessel is loaded with the correct equipment to perform a certain intervention effectively.

## 5.3. Punctuation: break-downs and anomalies

The strategies of LWI practices covered above involve getting the planning off the ground in the first place and subsequently making it more robust. Despite a month of planning, LWIs experience (small and not so small) anomalies. To facilitate a close link between LWI planning and execution, the intervention is always supervised by an LWI engineer onboard the vessel. The intervention is performed by representatives of the vendors of the subsea well equipment in question also onboard the vessel. Daily videoconferences are held between the LWI engineer onboard and those onshore. Usually the LWI engineer responsible for the planning of the given intervention is the one participating from the onshore end of these videoconferences.

That anomalies emerge following LWI planning is not surprising. LWI engineers work with objects they have hardly examined closely (physically) before. Their understandings, as described earlier, rely on vast historic and real-time data, consultations with both external partners and OGC colleagues.  Yet accuracy and completeness of information of wells varies, thus well interventions inherently involve risks.

Well engineers learn how to cope and improvise with aberrant wells over time. During an intervention, for instance, it is crucial to ensure that the equipment connected to the vessel is at the right depth. In one engineer's words, you need to know "where you are in the well". To ensure depth control, well engineers require detailed, updated information about the installed pipes. Lack of this information does not imply that an intervention cannot be performed:

> If we do not have certain information it means more uncertainty. If we do not know the length of pipes and how they are connected, we can do some workarounds. We can identify depth in several ways. We could use connection locator, which marks where a given pipe begins and were it ends. Alternatively, we could measure relative to the formation. So usually, we find a way out. (LWI engineer)

If depth control uncertainty arises, LWI engineers find alternative, compensating strategies to ensure accuracy. More significant anomalies arise, however, where such strategies are insufficient. The work-over equipment used to conduct the interventions are up to 20 meters in length and difficult to manoeuvre. During an intervention, the work-over equipment is manoeuvred via thin cables to the vessel. Well paths (i.e. the possibly several kilometres long trajectory of the well from the seabed to the reservoir) are hardly straight lines as they bend, even horizontally, to make them S-shaped. Despite precautions and preparations, incidents happen:

> It happens we drop 'things' [equipment] into the well. For instance, we can get stuck when we are going in or out of a well. In such case, we try to 'fish' the equipment up ourselves. However, we have limited possibilities from our vessel. If we cannot retrieve the equipment, we would, for instance, have to cut 4000 meters of cable plus the equipment attached to the cable. (LWI engineer)

Even 'straightforward' cases create dramatic anomalies:

> We made damages to a well during an intervention. We injected liquid into the well, but pumped too much, which subsequently increased the pressure causing the production tubing to collapse. We had to leave the well and a [mobile] rig had to be acquired to fix the well. (LWI engineer)

Coping with anomalies through learning over time about a specific well is necessary but not sufficient. Learning from anomalies *across* wells is also required. Depending on the level and type, an anomaly in one well is made visible in formal documents or designated arenas. With smaller anomalies, e.g. onboard LWI engineers relying on videoconferencing with onshore resources with knowledge of other wells:

> What do you do when you get stuck in a well? It is important to have someone [onshore LWI engineers] to discuss with. In extreme situations, offshore [LWI] engineers call the [LWI] engineer on duty who quickly assembles a support team. This team informs [governmental] authorities about the situations and contributes to offshore decision-making. You [the onboard LWI engineer] have to consider so many issues in these situations… (LWI engineer)

The so-called subsea pool is an institutional arena for discussing anomalies. It consists of well engineers from about 15 assets each with multiple wells from one geographical area of the Norwegian Continental Shelf. In their meetings, they discuss anomalies across wells with similar profile. Coping with such family resemblance of anomalies may lead to formalising new checklists or routines. To illustrate, in a sequence of meetings, recurring anomalies in multiple wells with the so-called work-over riser[12] were discussed. First, they decided to revise add an item to the existing checklist to check the riser specifically. Still not able to cope satisfactory, a later meeting decided on a protocol applicable to all wells in the assets for how the "riser shall be monitored with respect to fatigue life".

# 6. Discussion

## 6.1. Charactering the crafting of similarities

Our concern is more targeted than observing *that* there exist relationships between work practices. We analyse the specific relationship of similarity: *how* does similarity in work practices get crafted?

In a study of how Enterprise Systems acquire their similarity – their ability to be packaged to serve multiple client organisations – Pollock et al. (2007) argue that it is by 'generification'. Generification is interesting especially methodologically as it underscores the presence of arenas and actors more often than not left out in cases of Enterprise Systems implementations in a given organisation. The authors demonstrate how key actors (e.g. for Enterprise Systems: industry analysts like Gartner group) create a level of similarity as they operate across multiple clients. For LWI, the external service providers and vendors of subsea equipment in an analogous manner contribute towards similarities in LWI work practices. These external vendors serve multiple energy companies besides OGC. Their own ongoing efforts towards increased standardisation and interoperability of their equipment (with associated maintenance routines) rub off also to LWI practices.

Yamauchi and Swanson (2010) also analyse mechanisms through which similarities in technology mediated work practices. They suggest similarities in space emerge over time in 'familiarity pockets'. Due to time-space separation, the learning ('assimilation') implicated in establishing similar work practices is partial and thus limit the repertoire of work practices. Users, when challenged by anomalies or new situations, "rather than seek a deeper cognitive understanding…they tend to work around their ignorance" (ibid., p. 201). In contrast, the strategies we have identified for LWI practices definitely aim for 'deeper' understanding. What emerges from our analysis is an oscillating process. It starts with extracting the biography or 'personality' of the specific well. Document templates provide cues for which sources and whom to consult. Next, an implicit form of categorisation of the well is done – establishing its family resemblance - by identifying similar aspects with other wells. An important and institutionalised vehicle is here the filling in of a 2 x 2 risk matrix mapping frequency of incidents against consequences. LWI engineers are not concerned about directly comparing wells, only their risk profiles. Finally, there is the inevitable handling of anomalies where LWI planning meets the full, operational reality.

---

[12] A riser is a specialised tool used during LWI.

Coping with anomalies is not merely about in situ improvisations or accumulated learning from a specific well. This would effectively turn anomalies into a re-dressed version of practice theory. Crucial to our analysis is how anomalies from one well are made visible, thus potentially relevant, to other wells. Local anomalies feed an ongoing process of identifying family resemblance of anomalies across wells. This takes place through a combination of formal documents and institutionalised arenas/ meetings. Responses to anomalies are accordingly not handled (only) there and then, but get gradually sedimented into documents, checklists and routines. Significant anomalies need to be documented in a given system supervised by the National petroleum authority and are subject to nationally regulated audits. Smaller anomalies are documented in the Final well report. Patterns of similar anomalies are discussed in institutionalised arenas of well engineers from several assets and typically result in revised checklists and routines. LWI, especially for complex wells (high pressure and/ or temperature, history of incidents), have institutionalised 'peer review' processes as laid out by Governing documentation. A peer review process assembles a team of production and reservoir engineers to review documentation.

### 6.2. The pragmatism and politics of family resemblances

Starting from templates (uniform across space), LWI engineers develop a fragile, contingent, definitely fallible 'working knowledge' of the well, its personality and resemblance with related wells that allow similar enough LWI work practices. As the US Government appointed commission following the Deepwater Horizon blow-out in the Gulf of Mexico made painfully clear and LWI engineers would subscribe to,

"each oil well has its own personality" (2010, p.21). So if full knowledge about the local circumstances of a well is not the ambition, when and how do the engineers know sufficient for LWI to proceed? We discuss selected aspects.

LWI for one well never *is* the same as the next. LWI practices are made to be similar enough for ambulatory LWI to take place within technical, practical and institutional boundaries. This entails a *performative* rather than *essentialist* understanding of what similarity is (Orlikowski and Scott 2008). Similarity is not identifying certain attributes that all items share as observed by Rosch and Mervis (1975, p. 575) in their discussion of Wittgenstein's (1953) notion of family resemblance because "each item has at least one, and probably several, [attributes] in common with one or more other items, but *no, or few, [attributes] are common to all items*" (emphasis added).

Second, the question of similarity becomes a *pragmatic* one. Pragmatism has attracted some interest in IS research[13] but scholars have tended to pursue other aspects such as conceptual modelling (Ågerfalk, 2010) and action research (Sjöström & Goldkuhl, 2009). "Pragmatics", notes Giere (2004, p. 742) "has been largely a catchall for whatever is left over, but seldom systematically investigated". Rather than a fixed criterion, the crafting of similarity is inherently linked with the *intentionality/ purpose* of the work practice. The directedness of practical activities or in the words of Orr (1996, p. 6, emphasis in original), "The first and foremost goal of practice…is *getting the job done*", pragmatism shares with another crucial underpinning of practice theory, viz. phenomenology (Idhe 2001). For instance, in filling in the risk matrix for a well as part of the planning, the focus is to categorise the risk profile of the well, not capture its biography in full. This risk profile directs the type and extent of preparations for the LWI engineers 'to get the job done'.

Third, similarities in work practices are learnt over time (Chu and Robey 2008; Yamauchi and Swanson 2010). However, discussing similarities in work practices in terms of 'learning' downplays to the level of non-existence *political* aspects of attempting to establish similarities in work practices (Howard-Grenville and Carlile 2006). The vocabulary of learning is one purged of conflict. The pragmatic issue of similar enough work practices is, as Perin (2004) reminds us, caught in political cross-pressure from concerns of efficiency, safety and professional identity thus challenging the more harmonious, learning-oriented portray of e.g. high-reliability organisations (HRO) (Weick and Sutcliffe 2001). A vivid illustration of how safety is tied to concerns for similarities in work practices is the gas leakage problems OGC experienced at one of its fields in 2010. In the Governmental audit conducted after the incidents and subsequent shutdown, the main conclusion was that a disaster in the magnitude of Deepwater Horizon could have happened "under marginally different circumstances"[14]. The thrust of the audit's critique was OGC's apparently lacking ability to establish similar work practices for planning, production and maintenance across is oil and gas fields, including OGC's lacking ability to draw out the relevant similarities in wells separated in time and space. National petroleum authorities and OGC management want to move away from the traditional, local work practices to establish stronger similarities across space and time. In short, the ongoing efforts to establish more similar work practices across space-time is not only about 'learning from experience' – nobody opposes that – but a highly political issue involving the unions and management of OGC in addition to being framed within a national and international institutional and regulatory regime.

---

[13] See AIS Special Interest Group on Pragmatist IS Research (SIGPrag), http://www.sigprag.org/

# 7. Conclusion

In the context of processes of globalisation, Appadurai (1996) makes the observation that theorising lags significantly behind the empirically, unfolding phenomenon. This is not unlike the situation we are analysing. Business and public organisations have invested heavily to establish distributed yet uniform work practices, e.g. by introducing Enterprise Systems. As practice-based research makes clear, achieving identical ('best') work practices is unattainable. Still, had not managers, owners and investors after two decades of Enterprise Systems also recognised an interesting level of similarities in work practices, surely they would have fallen out of fashion? It seems to us that the unfolding, empirical phenomenon of technologically mediated efforts to promote similar work practices has yet to receive an adequate theoretical account in information systems research.

In our use the notion of family resemblance is *performative*, *pragmatic* and *political*. Resonating with Wittgenstein's (1953) original insights, family resemblance of work practices is not about sharing certain attributes. Similarity is *performed* or crafted through the strategies we have discussed. The criterion for when sufficient similarity is achieved is pragmatic in the sense of directed or intentional. The filling in of a 2 x 2 risk matrix directs attention to those similarities that matter for the planning of the LWI (e.g. type of equipment to bring along). Moreover, the crafting of similarity is *political*. This is especially evident around issues of safety and risk. Whether compliance to uniform work practices improves safety, or whether safe operations and maintenance are better served by practices shaped by the local circumstances of the well, is discussed heatedly between OGC management and unions as well as national authorities.

---

[14] www.ptil.no, the Petroleum Safety Authorities, Norway.

Family resemblance of work practices is relevant to many organisations for reasons of economic performance, quality of service and safety to human life and the environment. A key practical implication from our study is that the crafting of family resemblance is ongoing and emergent rather than about nailing the exact balance between uniformity and localised work practices. More specifically, our study show the importance of flexibly *stepping up* the degree of formalism and amount of resources in response to number, frequency and type of anomaly. As anomalies increase in gravity, so do the number of people, amount of time spent, format for deliberation/ arena and degree of formal documentation. Rather than a fixed, institutional response, a dynamically modulated response relative to the gravity of the anomalies is required.

**Acknowledgement**

We have benefited from interactions with many including: Petter Almklov, Bendik Bygstad, Gunnar Ellingsen, Ole Hanseth, Torgeir Haavik, Ola Henfridsson, Neil Pollock, Ola Titlestad, Margunn Aanestad. Editor Dan Robey and our two anonymous reviewers have consistently provided us with challenging and constructive comments. Our work has in part been sponsored by a grant by the Norwegian Research Council's Verdikt programme.

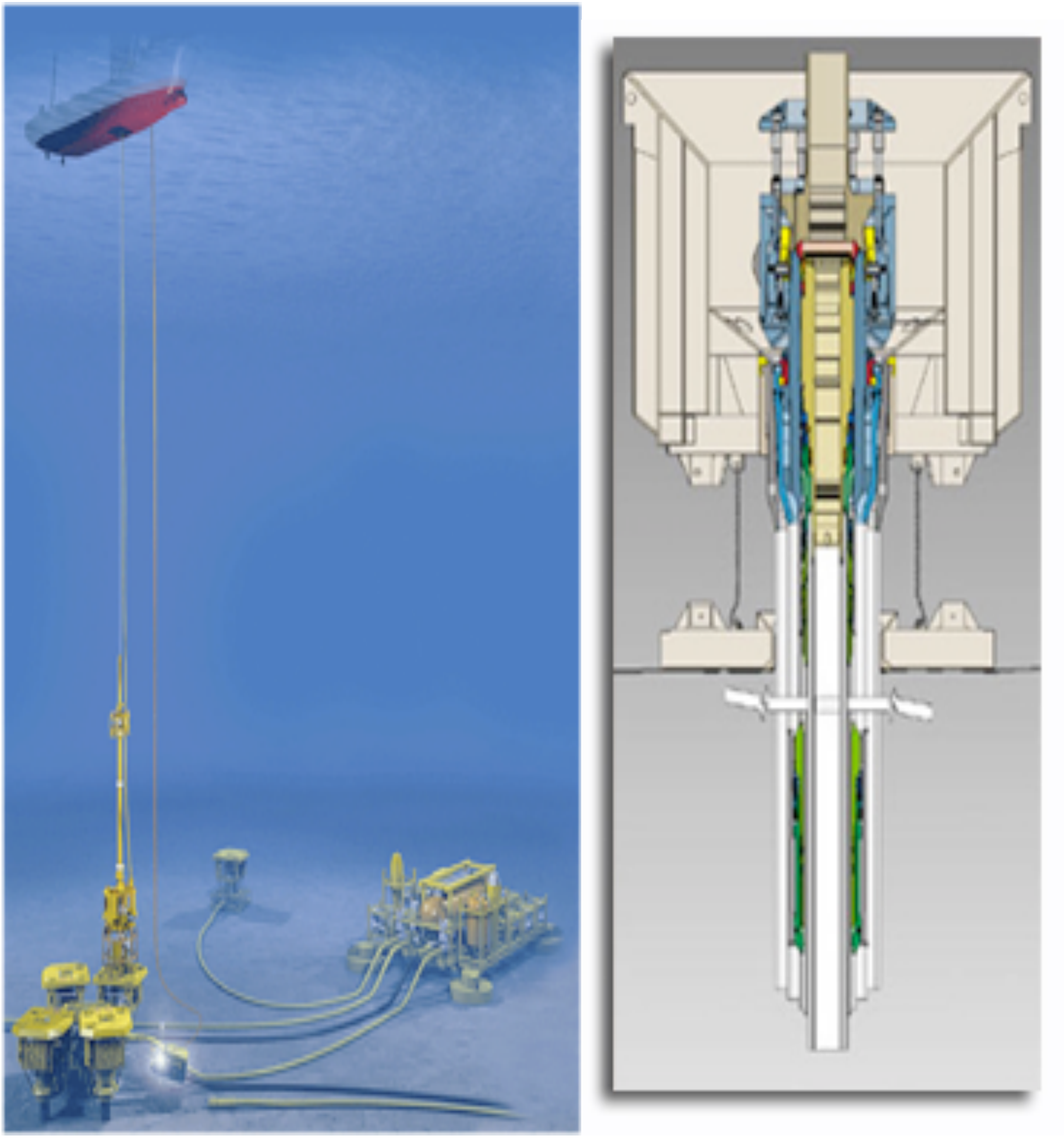

Figure 1. Illustration of subsea equipment operated by a vessel (left) and a subsea wellhead (right).

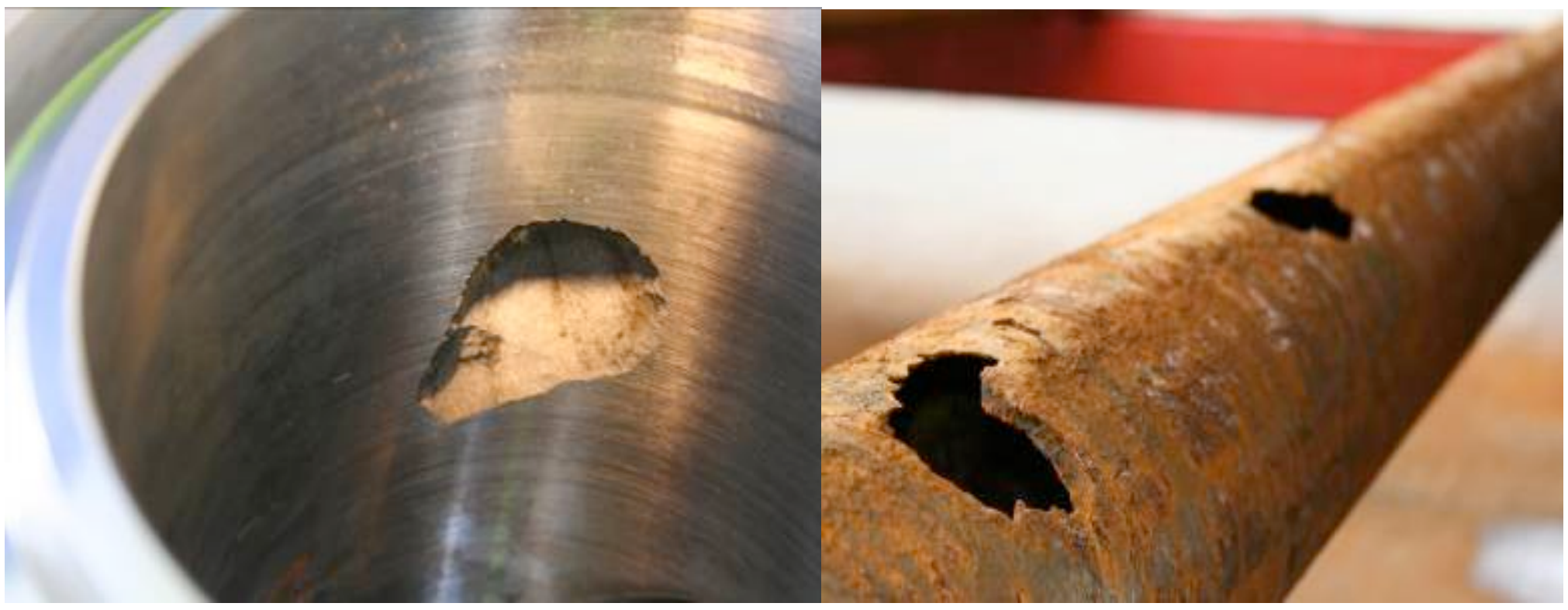

Figure 2. Illustrations of corrosion in of a well’s tubing.

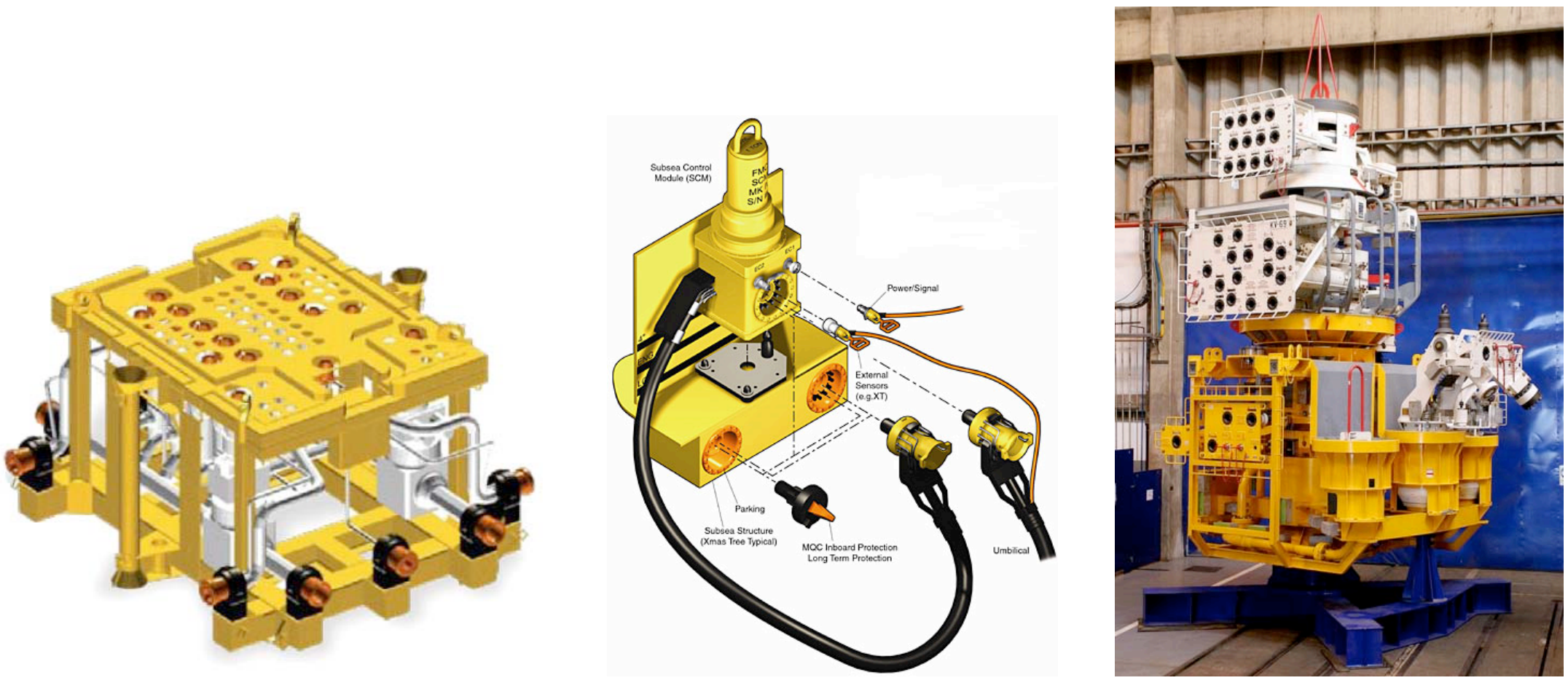

Figure 3. Illustrations of key subsea components: manifold (left), control module (middle) and x-mas tree (right).

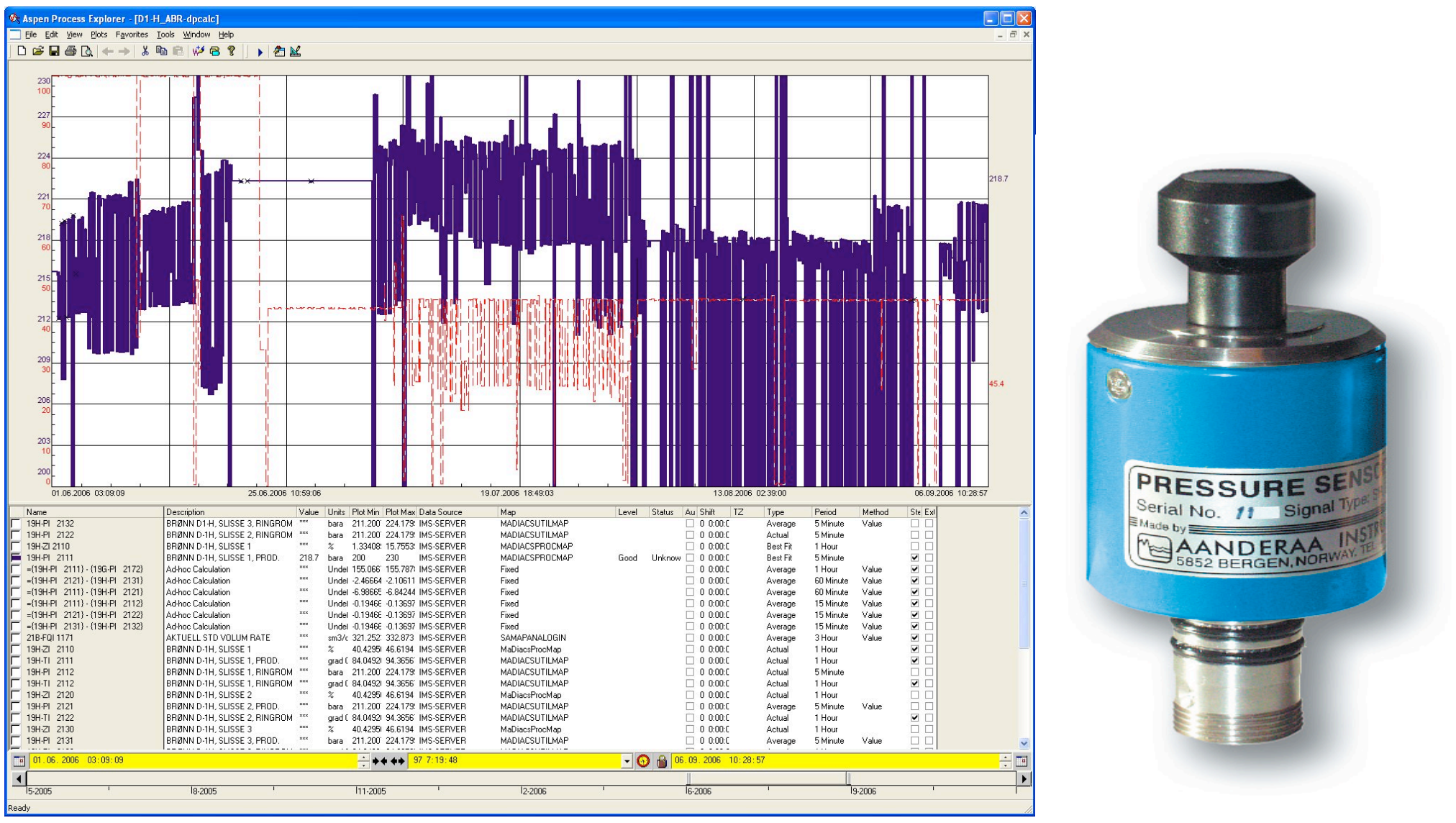

Figure 4. Readings (left) from the pressure sensor (right) in the form of a histogram over time.

| Actor group/ department | Number of interviews |
|---|---|
| IT-managers and developers | 14 |
| Research & Development department (R&D) | 23 |
| Oil and Gas Production (OGP) | 22 |
| Light well intervention (LWI) department | 9 |
| Total number of interviews | 68 |

Table 1. Overview over type and number of interview informants.

| Construct | Conceptual category | Excerpt |
|---|---|---|
| Differentiation | Biography | "Every well is unique" |
| | | "You need to know the personality of the well" |
| | | "If you didn't follow the well from its inception, there is no way you know where to find the information [about the well]" |
| | Configuration | "There is no plug & play possibilities [across vendors of subsea equipment]" |
| | | "The technical complexity of subsea equipment is challenging" |
| | | "I call the vendor for technical details [of equipment]" |
| Assembling similarities | Triangulate | "Information in [name of system] [need to be] compared with other sources, e.g. [name of system]" |
| | | "[those] sensors stopped working months ago" |
| | Trust | "He [referring to a collegue] can identify mistakes because he has worked with wells for 15 years" |
| | | "If I lack specific information…I call an operator [at the platform's control room]" |
| Punctuation | Uncertainty | "If I…feel uncertain…[about] certain limitations in the well, the platform [operator] will know more about these" |
| | | The Safe Job Assessment requires filling in the 2 x 2 risk matrix, plotting frequency of incidents against consequences. |
| | | "If we do not know the length of the pipes and how they are connected, we can do some workarounds" |
| | Unplanned events | "Recently we discovered that the control system of the x-mas tree [cf. Figure 2] was not compatible with our equipment" |
| | | "It happens we drop 'things' [equipment] into the well" |
| | | "We injected liquid into the well, but pumped too much…causing the production tubing to collapse" |

Table 2. Interpretative constructs, conceptual categories and excerpts underpinning our data analysis.